\documentclass[12pt]{article}

\begin{document}

\begin{titlepage}

\begin{flushright}
IUHET-487\\
\end{flushright}
\vskip 2.5cm

\begin{center}
{\Large \bf Lorentz Violation and Faddeev-Popov Ghosts}
\end{center}

\vspace{1ex}

\begin{center}
{\large B. Altschul\footnote{{\tt baltschu@indiana.edu}}}

\vspace{5mm}
{\sl Department of Physics} \\
{\sl Indiana University} \\
{\sl Bloomington, IN 47405 USA} \\

\end{center}

\vspace{2.5ex}

\medskip

\centerline {\bf Abstract}

\bigskip

We consider how Lorentz-violating interactions in the Faddeev-Popov ghost sector
will affect scalar QED. The behavior depends sensitively on whether the gauge
symmetry is spontaneously broken. If the symmetry is not broken, Lorentz
violations in the ghost sector are unphysical, but if there is spontaneous
breaking, radiative corrections will induce Lorentz-violating and gauge-dependent
terms in other sectors of the theory.

\bigskip

\end{titlepage}

\newpage

\section{Introduction}

Recently there has been quite a bit of interest in the suggestion that Lorentz
symmetry may not be exact in nature. Small violations of this fundamental
symmetry could arise in connection with the novel physics of the Planck scale.
One major focus of research has been the embedding of possible Lorentz-violating
effects in effective field theories.
The general local Lorentz-violating standard model extension (SME) has been
developed~\cite{ref-kost1,ref-kost2,ref-kost12}, and the
stability~\cite{ref-kost3} and
renormalizability~\cite{ref-kost4} of this extension have been studied. The
general SME contains all possible local operators that may be constructed out
of existing standard model fields. However, typically one will consider only 
a more limited
subcollection of these operators, such as the minimal SME, which contains only
superficially renormalizable operators that are invariant under the standard
model gauge group.

The minimal SME provides an excellent framework
within which to analyze
the results of experimental tests of special relativity.
To date, such
experimental tests have included studies of matter-antimatter asymmetries for
trapped charged particles~\cite{ref-bluhm1,ref-bluhm2,ref-gabirelse,
ref-dehmelt1} and bound state systems~\cite{ref-bluhm3,ref-phillips},
determinations of muon properties~\cite{ref-kost8,ref-hughes}, analyses of
the behavior of spin-polarized matter~\cite{ref-kost9,ref-heckel},
frequency standard comparisons~\cite{ref-berglund,ref-kost6,ref-bear},
measurements of neutral meson
oscillations~\cite{ref-kost10,ref-kost7,ref-hsiung,ref-abe},
polarization measurements on the light from distant galaxies~\cite{ref-carroll1,
ref-carroll2,ref-kost11}, and others.

There can be a very subtle interplay between Lorentz
violation and gauge invariance. For example, a Lorentz-violating Chern-Simons term
${\cal L}_{CS}=\frac{1}{2}k_{\mu}\epsilon^{\mu\alpha\beta\gamma}F_{\alpha\beta}
A_{\gamma}$~\cite{ref-jackiw2,ref-schonfeld,ref-carroll1} in the Lagrange density
is not gauge invariant. However, since ${\cal L}_{CS}$
changes by a total derivative
under a gauge transformation, the integrated action is gauge invariant, and the
equations of motion involve only the field strength $F^{\mu\nu}$.
The quantum corrections to ${\cal L}_{CS}$ in spinor QED are even more
complicated, and what kind of gauge invariance the final theory possesses depends
sensitively on how the theory is regulated~\cite{ref-coleman1,ref-jackiw1,
ref-victoria1,ref-chen,ref-chung3,ref-altschul1,ref-altschul2}. In particular, the
radiatively-induced Chern-Simons term is necessarily finite, but its value is
not uniquely determined.

In this paper, we shall examine some further properties of Lorentz-violating
quantum field theory. The focus will be on quantum corrections, particularly
those associated with the Faddeev-Popov ghosts that arise in the quantization
of gauge theories. Since the presence of these ghosts is subtly entwined with the
symmetry properties of these theories, we expect that any changes to the ghosts
sector's structure could have a significant impact on gauge invariance. In this
section, we shall review the structure of the gauge-fixed scalar QED Lagrangian.
In section~\ref{sec-calc}, we shall introduce Lorentz-violating modifications to
this gauge theory and calculate the scalar field self-energy in their presence.
The physical interpretation of our results is discussed in
section~\ref{sec-discuss}, and our conclusions summarized in
section~\ref{sec-conclusion}.

Our starting point will be the gauge-fixed Faddeev-Popov Lagrange density for
scalar QED (in a generic $R_{\xi}$ gauge),
\begin{equation}
\label{eq-L}
{\cal L}=-\frac{1}{4}F^{\mu\nu}F_{\mu\nu}+\left|D^{\mu}\Phi\right|^{2}-V(\Phi)
-\frac{1}{2\xi}\left(\partial^{\mu}A_{\mu}-\xi ev\varphi\right)^{2}+\bar{c}
\left[-\partial^{2}-\xi m_{A}^{2}\left(1+\frac{h}{v}\right)\right]c.
\end{equation}
$F^{\mu\nu}=\partial^{\mu}A^{\nu}-\partial^{\nu}A^{\mu}$ is the Abelian field
strength, $D^{\mu}=\partial^{\mu}+ieA^{\mu}$ is the covariant derivative, and
$V(\Phi)$ is
the scalar field potential. The complex scalar field $\Phi$ itself we parameterize
as
\begin{equation}
\Phi(x)=\frac{1}{\sqrt{2}}[v+h(x)+i\varphi(x)].
\end{equation}
The potential might or might not induce spontaneous symmetry breaking, and $v$
is the (possibly vanishing) vacuum expectation value of the field, which we have
taken to be in
the real direction. The gauge fixing terms then depend on the fields $h$ and
$\varphi$; if $v\neq 0$, these
are the Higgs and Goldstone boson fields, respectively. The parameter $\xi$
determines the choice of gauge, and in the absence of explicit symmetry breaking
terms, it should cancel out in all physical results. Finally, $m^{2}_{A}=e^{2}
v^{2}$ is the mass of the physical gauge field, and $c$ and $\bar{c}$ are
Grassmann-valued ghost fields.

The potential $V(\Phi)$ may be left fairly general. We shall not make use of
any of its properties, except the value of $v$ it induces. So $V(\Phi)$ need not
be the
bare scalar potential of the theory. If the potential possesses a
symmetry-breaking minimum, that minimum
could be the result of radiative corrections (as in the Coleman-Weinberg
model~\cite{ref-coleman2}) or strong interactions in the matter sector.

The $R_{\xi}$ gauge Lagrange density (\ref{eq-L}) may be obtained by the
Feddeev-Popov procedure~\cite{ref-feddeev}. We begin with the conventional scalar
QED
Lagrange density,
\begin{equation}
{\cal L}_{0}=-\frac{1}{4}F^{\mu\nu}F_{\mu\nu}+\left|D^{\mu}\Phi\right|^{2}-
V(\Phi).
\end{equation}
Then we introduce a gauge fixing function
\begin{equation}
G=\frac{1}{\sqrt{\xi}}\left(\partial^{\mu}A_{\mu}-\xi ev\varphi\right).
\end{equation}
Following standard procedure, we integrate over different values of $G$, weighted
by a Gaussian. This transforms the Lagrange density to ${\cal L}_{0}-\frac{1}{2}
G^{2}$.

To complete the Faddeev-Popov quantization procedure however, we must also
include the ghosts. The ghost Lagrangian is determined by the gauge variation of
$G$. That gauge variation is represented by the determinant of the operator
$\delta G/\delta\alpha$,
where $\alpha(x)$ is the parameter of a local gauge transformation,
\begin{eqnarray}
\delta h & = & -\alpha\varphi \\
\delta\varphi & = & \alpha (v+h) \\
\delta A^{\mu} & = & -\frac{1}{e}\partial^{\mu}\alpha.
\end{eqnarray}
Since
\begin{equation}
\frac{\delta G}{\delta\alpha}=\frac{1}{\sqrt{\xi}}\left[-\frac{1}{e}\partial^{2}-
\xi m_{A}(v+h)\right],
\end{equation}
the ghost term in (\ref{eq-L}) reproduces the functional determinant of $\delta G
/\delta\alpha$ when it is integrated out. This term therefore completes the
gauge-fixed $R_{\xi}$ Lagrange density.

In the conventional framework that we have outlined,
the ghosts $c$ and $\bar{c}$ are seen as auxiliary fields; they are
introduced as part of the gauge-fixing procedure, the purpose of which is to
reorganize the action, so that the zero modes of the gauge action do not
interfere with the derivation of the propagator. So the Faddeev-Popov ghosts can
be seen as further manifestations of the fundamental gauge field; in addition to
the gauge-fixed $A^{\mu}$, the gauge sector contains these anticommuting fields.

However, another slightly different viewpoint is also possible. Since (\ref{eq-L})
provides a correct and complete description of scalar QED, we could take this
gauge-fixed action as our basic description of the physics.
Then if we are interested in describing all possible Lorentz-violating
modifications
of scalar QED, we should include those Lorentz-violating operators
that involve ghosts.

The Lagrange density (\ref{eq-L}), since it is gauge fixed, does not possess a
conventional $U(1)$ local symmetry. So we must be a little careful when we speak
of the gauge invariance of this Lagrangian. However,
since (\ref{eq-L}) can be derived
from a truly gauge invariant expression by the Faddeev-Popov
procedure, this
Lagrange density does have gauge invariance in a certain sense. Yet because
the gauge symmetry is somewhat obscured, it may not necessarily be clear whether
a particular term, if added to ${\cal L}$, will break the symmetry or not.
One might hope that this difficulty could be resolved by examining the BRST
symmetry of the gauge-fixed Lagrangian~\cite{ref-becchi,ref-tyutin}. However, this
turns out not to suffice. We shall encounter operators which break BRST symmetry
when added to (\ref{eq-L}), yet which do not change the fact that the physical
theory one would observe is an Abelian gauge theory.

Before we introduce Lorentz violation and begin calculating loop diagrams, we
should point out one further point.
If the gauge symmetry is not spontaneously broken, then the gauge-fixed
Lagrange density reduces to
\begin{equation}
{\cal L}=-\frac{1}{4}F^{\mu\nu}F_{\mu\nu}+\left|D^{\mu}\Phi\right|^{2}-V(\Phi)
-\frac{1}{2\xi}\left(\partial^{\mu}A_{\mu}\right)^{2}+(\partial^{\mu}\bar{c})
(\partial_{\mu}c).
\end{equation}
What is important about this expression is the well-known fact that the
Fedeev-Popov ghosts decouple completely. They can be completely ignored, and
even if they are included in the theory's Feynman diagrams, they will only appear
in unconnected vacuum bubbles. So the structure of the ghost sector has no effect
on the $S$-matrix. This is in sharp contrast with the case in which
the gauge symmetry is spontaneously broken, because if $v$ is nonzero,
(\ref{eq-L}) implies that there is a coupling between the ghosts and the
physical Higgs field $h$. It is this difference that will be at the crux of our
discussions.

\section{Lorentz Violation from the Ghost Sector}
\label{sec-calc}

\subsection{Lorentz-Violating Ghost Lagrangian}

We shall now consider modifying (\ref{eq-L}) to include Lorentz violation in the
ghost sector.
However, not all the Lorentz-violating terms that we may add to ${\cal L}$
are physically meaningful. There are actually relatively few superficially
renormalizable couplings one can write down involving only the $c$ and $\bar{c}$
fields, because these fields are Lorentz scalars. There is only
one such CPT-odd modification of the ghost sector; adding it changes the Lagrange
density for $c$ and $\bar{c}$ into
\begin{equation}
{\cal L}_{a}=\left[\left(\partial^{\mu}+ia^{\mu}\right)\bar{c}\right]
\left[\left(\partial^{\mu}-ia^{\mu}\right)c\right]
-\xi m_{A}^{2}\left(1+\frac{h}{v}\right)\bar{c}c.
\end{equation}
However, the presence of $a^{\mu}$ actually has no physical consequences.
A field redefinition
\begin{equation}
\label{eq-atrans}
c\rightarrow e^{ia\cdot x}c,\,\bar{c}\rightarrow e^{-ia\cdot x}\bar{c}
\end{equation}
eliminates $a^{\mu}$ from the theory. This shift is equivalent to a change in the
origin of the momentum integration for all ghost loops. In more general theories,
field redefinitions may be used to eliminate a number of other apparently
Lorentz-violating terms~\cite{ref-colladay2}.

A superficially renormalizable, CPT-even modification of the ghost sector is also
possible. In this case, the Lorentz violation changes the ghost Lagrange density
to
\begin{equation}
\label{eq-Lc}
{\cal L}_{c}=\bar{c}\left[-\partial^{2}-c^{\nu\mu}\partial_{\nu}\partial_{\mu}
-\xi m_{A}^{2}\left(1+\frac{h}{v}\right)\right]c.
\end{equation}
This does not represent a Lorentz-violating choice of gauge; it is
something entirely different. Using a Lorentz-violating gauge would mean
choosing a function $G$ that transforms nontrivially under particle Lorentz
transformations. Doing this would induce Lorentz violation in the $A^{\mu}$
sector, which any violations in the ghost sector should then cancel out. Here,
we have added Lorentz violation to the ghost Lagrangian without making any
corresponding changes to the Lagrangian for $A^{\mu}$.

The main question that this paper shall address is whether the Lorentz-violating
coefficient $c^{\nu\mu}$ is physical. It is obvious from the form of the
interaction that the antisymmetric part of $c^{\nu\mu}$ does not contribute.
However, whether the symmetric part will manifest itself physically is not so
obvious. In fact, we shall show that the question of whether the
$c^{\nu\mu}$ Lorentz violation contributes to real effects cannot be answered
by looking at the gauge sector alone. The behavior of the matter fields affects
things in a crucial way.

{\em Ad hoc} modifications of the ghost sector like (\ref{eq-Lc})
will also be expected to damage
the gauge invariance properties of our theory. However, if $c^{\nu\mu}$ turns
out to be unphysical (like $a^{\mu}$), then the gauge symmetry is effectively
restored. If the Lorentz-violating interactions do not contribute to physical
effects, then they may be ignored, and only the gauge-symmetric part of the
theory need be retained. So in this sense, the antisymmetric part of $c^{\nu\mu}$
does not violate gauge invariance, just as it does not violate physical Lorentz
invariance.

\subsection{Unbroken $U(1)$ Phase}

Since the ghosts are not supposed to appear as external particles, addressing
the issues we wish to discuss
must necessarily involve consideration of quantum corrections.
The Feynman rules for the $c^{\nu\mu}$-modified theory depend on whether the
scalar field potential induces spontaneous symmetry breaking. If it does not,
then $v$ and $m_{A}$ vanish, so the Faddeev-Popov ghosts remain decoupled from
the rest of the theory. The presence of the Lorentz violation does not change
this. So $c$ and $\bar{c}$ still only appear in vacuum bubble diagrams. These
disconnected diagrams will be modified, but this fact does not have any physical
meaning (since the theory as we are considering it is not coupled to gravity).
We may therefore conclude that $c^{\nu\mu}$ does not correspond to any physical
Lorentz violation, as long as the theory is in a phase with no spontaneous
breaking of the $U(1)$ symmetry. Nor is there
any meaningful breaking of gauge invariance under these circumstances
(even though the addition of $c^{\nu\mu}$ destroys the BRST symmetry).

\subsection{Broken $U(1)$ Phase}

The case in which the gauge symmetry is broken
by a nonzero $v$ is much more complicated, because
the ghosts do not decouple. Instead, they are coupled to the physical Higgs field
$h$. The ghosts will contribute to the $n$-point correlation functions for $h$;
the important diagrams have $n$ Higgs lines attached to a ghost loop. We shall
compute the simplest such diagrams, which give a one-loop contribution to the
Higgs two-point function. (The one-loop diagrams with $n>2$ external lines are all
finite by power counting.)

We shall work to leading order in $c^{\nu\mu}$. It is common practice to ignore
higer-order Lorentz-violating effects, because any physical Lorentz violation is
known to be small.
At first order in $c^{\mu\nu}$, there are two
diagrams, which we shall evaluate by dimensional regularization.
Without Lorentz violation, the one-loop ghost contribution to the Higgs
self-energy $\Pi_{h}(p)$ comes from a single diagram with two ghost propagators.
The leading
Lorentz-violating contributions then come from adding $c^{\nu\mu}$ insertions
on one or the other of the ghost lines. This splits the ${\cal O}(c^{\nu\mu})$
contribution into two terms, $\Pi_{h}^{1}$ and $\Pi_{h}^{2}$, and the
first diagram of this type gives
\begin{equation}
i\Pi_{h}^{1}(p)=(-1)\left(-i\xi\frac{m_{A}^{2}}{v}\right)^{2}
\int\frac{d^{d}k}{(2\pi)
^{d}}\,\frac{i}{k^{2}-\xi m_{A}^{2}}\left(-ic^{\nu\mu}k_{\nu}k_{\mu}\right)
\frac{i}{k^{2}-\xi m_{A}^{2}}\frac{i}{(k+p)^{2}-\xi m_{A}^{2}},
\end{equation}
where $p$ is the external momentum of the $h$ field. The factor of $-1$ comes
from the Grassmann nature of the ghosts. This whole expression depends
strongly on the gauge parameter $\xi$, and at $\xi=0$ it vanishes. This dependence
just indicates that if we find a nonzero physical result, it will break gauge as
well as Lorentz symmetry, just as previously discussed.

We may combine the denominators with a Feynman parameter $x$ to get
\begin{equation}
i\Pi_{h}^{1}(p)=-\xi^{2}\frac{m_{A}^{4}}{v^{2}}\int_{0}^{1}dx\int\frac{d^{d}k}
{(2\pi)
^{d}}\,c^{\nu\mu}p_{\nu}p_{\mu}\frac{2(1-x)}{\left\{(1-x)(k^{2}-\xi m_{A}^{2})+
x\left[(k+p)^{2}-\xi m_{A}^{2}\right]\right\}^{3}}.
\end{equation}
In terms of $\ell=k+xp$ and $\Delta=-x(1-x)p^{2}+\xi m_{A}^{2}$, and dropping
all terms odd in $\ell$, this is
\begin{equation}
i\Pi_{h}^{1}(p)=-2\xi^{2}\frac{m_{A}^{4}}{v^{2}}\int_{0}^{1}dx\int\frac{d^{d}k}
{(2\pi)
^{d}}\,c^{\nu\mu}\frac{(1-x)(\ell_{\nu}\ell_{\mu}+x^{2}p_{\nu}p_{\mu})}
{(\ell^{2}-\Delta)^{3}}.
\end{equation}
The $\ell_{\nu}\ell_{\mu}$ term makes a contribution proportional to $g_{\nu\mu}$.
When contracted with $c^{\nu\mu}$, this gives only a Lorentz scalar. The
momentum-independent part of this becomes
part of the mass renormalization and does not result in any Higgs sector Lorentz
violation. Moreover, despite its dependence on $\xi$, the $p$-independent term
cannot result in
a physical failure of gauge invariance, because the measurable value of the Higgs
mass is a free parameter in the renormalized theory. We may absorb the dependence
on $\xi$ into the unphysical bare mass of the Higgs. We shall
therefore not consider the
divergent part of this expression any further, although we shall comment on the
$p^{2}$-dependent part of
the Lorentz-invariant term at the end of this section.

The $p_{\nu}p_{\mu}$ term gives the potentially Lorentz-violating part of
$\Pi_{h}^{1}(p)$, which we shall denote as $\Pi_{h,LV}^{1}(p)$. Evaluating this
as $d\rightarrow4$, we find
\begin{equation}
i\Pi_{h,LV}^{1}(p)=\frac{i}{16\pi^{2}}\xi^{2}\frac{m_{A}^{4}}{v^{2}}\int_{0}^{1}
dx\,\frac{x^{2}(1-x)}{\xi m_{A}^{2}-x(1-x)p^{2}}c^{\nu\mu}p_{\nu}p_{\mu}.
\end{equation}
The Lorentz-violating contribution $\Pi_{h,LV}^{2}(p)$ coming from the other
diagram is the same, but with $x\rightarrow(1-x)$. So adding these together,
we get
\begin{equation}
i\Pi_{h,LV}(p)=\frac{i}{16\pi^{2}}\xi^{2}\frac{m_{A}^{4}}{v^{2}}\int_{0}^{1}dx\,
\frac{x(1-x)}{\xi m_{A}^{2}-x(1-x)p^{2}-i\eta}c^{\nu\mu}p_{\nu}p_{\mu},
\end{equation}
where we have inserted an infitesimal $i\eta$ ($\eta>0$) to give the correct
behavior when $\Delta$ vanishes.
If $0<p^{2}<4\xi m_{A}^{2}$, the integral over $x$ may be evaluated
in closed form:
\begin{equation}
\int_{0}^{1}dx\,\frac{x(1-x)}{\xi m_{A}^{2}-x(1-x)p^{2}}=-\frac{1}{p^{2}}+\frac{
4\xi m_{A}^{2}}{(p^{2})^{3/2}\sqrt{4\xi m_{A}^{2}-p^{2}}}\tan^{-1}
\left(\frac{\sqrt{p^2}}{\sqrt{4\xi m_{A}^{2}-p^{2}}}\right).
\end{equation}

We can also combine the Lorentz-invariant parts of $\Pi_{h}^{1}(p)$ and
$\Pi_{h}^{2}(p)$ (coming from the $\ell_{\nu}\ell_{\mu}$ terms), to obtain another
expression, $\Pi_{h,LI}(p)$. As previously stated, the divergent,
$p^{2}$-independent part of $\Pi_{h,LI}(p)$ is just absorbed
into the mass renormalization. However, there is a finite, momentum-dependent part
as well. The momentum dependence is given by
\begin{equation}
i\Pi_{h,LI}(p)=-\frac{i}{32\pi^{2}}\xi^{2}\frac{m_{A}^{4}}{v^{2}}c^{\mu}\,_{\mu}
\int_{0}^{1}dx\, \log[\xi m_{A}^{2}-x(1-x)p^{2}-i\eta]+C.
\end{equation}
$C$ is the unphysical infinite constant (which, however, contains the scale of
the logarithm). Although the momentum-dependent part of
$\Pi_{h,LI}(p)$ is Lorentz invariant, it is finite, and most of the remarks
in section~\ref{ssec-broken} will apply to this expression as well as to
$\Pi_{h,LV}(p)$.

\section{Interpretation of Results}
\label{sec-discuss}

\subsection{Unbroken Phase}

We shall now look at the physical implications of these results, beginning in the
phase without spontaneous symmetry breaking. Of course, matters are very simple
in this situation, because we have already established that there can be no
contribution from the ghosts to any connected diagram with physical external
particles.

The physical theory remains an Abelian gauge theory. The $c^{\nu\mu}$ modification
of the ghost sector, although it apparently violates Lorentz and BRST symmetries,
does not affect the symmetries of any physical process. Indeed, it does not
affect physical processes in any way. The $S$-matrix for the theory is unchanged
by the modification of the Lagrangian.

It is even possible to see how $c^{\nu\mu}$ could be defined away in the path
integral. However, this is not accomplished by a field redefinition, but rather
by replacing the existing ghosts with an entirely new set of ghost fields.
$c$ and $\bar{c}$ can simply by integrated out of the theory; this will
only change the normalization of the measure. Then a new set of ghost fields $c'$
and $\bar{c}'$ may be introduced, with a rescaled functional measure and
a Lagrange density $(\partial^{\mu}\bar{c}')(\partial_{\mu}c')$. This restores
the Lagrangian to its Lorentz-invariant form. Of course, all these formal
manipulations are rather trivial, but this just underscores the fact that the
ghosts are entirely superfluous in this theory.

\subsection{Broken Phase}
\label{ssec-broken}

Things are not trivial in the Higgs phase, however.
Unless $\xi=0$, $\Pi_{h,LV}(p)$ makes a physical Lorentz-violating contribution to
the Higgs
propagator. Let us examine a few special cases in the parameter space
and see how this radiative correction will affect
particle propagation. For an on-shell Higgs boson, $p^{2}=m_{h}^{2}$, where
$m_{h}$ is the Higgs mass. Then, if $|\xi m_{A}^{2}|\gg m_{h}^{2}$, we have
\begin{equation}
\Pi_{h,LV}(p^{2}=m_{h}^{2})=\frac{1}{96\pi^{2}}\xi\frac{m_{A}^{2}}{v^{2}}
c^{\nu\mu}p_{\nu}p_{\mu}.
\end{equation}
In the opposite, limit $|\xi m_{A}^{2}|\ll m_{h}^{2}$, the contribution to
the on-shell propagator is given by
\begin{equation}
\Pi_{h,LV}(p^{2}=m_{h}^{2})=-\frac{1}{16\pi^{2}}\xi^{2}\frac{m_{A}^{4}}
{v^{2}m_{h}^{2}}c^{\nu\mu}p_{\nu}p_{\mu}.
\end{equation}
(Note that the relationship between $\xi m_{A}^{2}$ and $m_{h}^{2}$ is not
determined merely by the ordinarily physical masses.
The relative sizes of these terms depend on the normally unphysical gauge
parameter $\xi$.)
In either limit, there is a nonzero Lorentz-violating term in the effective action
for the Higgs. This term will affect the propagation states of the theory, just
as would a Lorentz-violating tensor appearing in the fundamental Lagrangian. The
particular terms that we have found would, for example, change the energy-momentum
relation and hence velocity for physical Higgs excitations~\cite{ref-altschul3}.

So in the Higgs phase, when the $U(1)$ gauge symmetry is
spontaneously broken, the symmetric part of $c^{\nu\mu}$ has a real physical
effect. The magnitude of the induced violation in the Higgs sector
is controlled by $\xi$. In the
Lorenz-Landau gauge ($\xi=0$), $\Pi_{h,LV}(p)$ actually vanishes; but
otherwise, the Lorentz violation will be nonzero. This
explicit $\xi$ dependence signals that there is also a breakdown of gauge
invariance. The Higgs propagator has acquired a new gauge dependence, which will
not be cancelled by effects in other sectors of the theory.

However, although $\Pi_{h,LV}$ and $\Pi_{h,LI}$ both depend on $p$, the momentum
of the Higgs particle, rather than $p-eA$, this should not be
taken as a further indication of gauge invariance violation. We have
not considered any diagrams with external photons, and in
fact, when such higher-order terms
are included, the proper dependence on covariant derivatives should appear.

The induced Lorentz violation in the unitarity gauge cannot be studied by taking
the $\xi\rightarrow\infty$ limit of $\Pi_{h,LV}(p)$. That limit
would have to be taken
before any loop integrations are performed~\cite{ref-dolan}. This limiting
process will cause any diagrams containing more ghost propagators than Higgs-ghost
vertices to vanish. All diagrams involving $c^{\nu\mu}$ have this property, so
there is again no Lorentz violation in this singular limit. However, this is
fairly unsurprising, since for $\xi=\infty$, the propagating ghosts
effectively disappear.

What we have found is a somewhat unexpected connection between the gauge sector
of the theory (of which the ghosts are a part) and the Higgs potential.
Some aspects of
the entanglement between the gauge and Higgs sectors in spontaneously broken gauge
theories are already well known. The most obvious example is the ``eating'' of the
Goldstone boson by the gauge field; the fundamental scalar becomes the
longitudinal component of the vector boson.
We can see in (\ref{eq-L}) where information about the Higgs potential has been
fed back into the gauge sector. The ghost Lagrangian depends on $v$, which is not
a quantity that can be determined within the gauge sector. However, we should
keep in mind that (\ref{eq-L}) is valid in either the Higgs or unbroken phase
of the theory; the appearance of $v$ in the ghost Lagrangian is not alone
responsible for the difference in behavior between the two phases of the theory.

In actuality, the theory may not even be well-defined in the broken phase. The
gauge-dependent corrections could destroy renormalizability, and without BRST
symmetry, we cannot necessarily ensure the unitarity of the $S$-matrix. However,
we cannot know whether either of these two problems actually exists without
performing more detailed calculations. Gauge invariance is not actually a
necessary requirement for the renormalizability of an Abelian vector
theory, and although the BRST symmetry is also broken in the $v=0$ case, we know
that the resulting theory is definitely unitary. Moreover, we can be certain that
if $\xi=0$, there will be a well-defined theory, because all the troublesome
radiative corrections vanish.

\subsection{Relationship to Finite, Undetermined Quantum Corrections}

The fact that the finite radiative corrections are not gauge invariant
in the Higgs phase allows us to
draw an analogy with other finite, yet undetermined, radiative corrections.
The parameter $\xi$ which controls the size of the radiative corrections does not
appear in any tree-level results. Although the Feynman rules depend on its value,
this dependence cancels in all physical quantities. However, once loop corrections
are included, Lorentz-violating effects can appear. The size of these effects
depends on $\xi$, which is effectively a free parameter. That is, the loop
corrections cannot be
uniquely determined from observations of tree-level processes.

A theoretical discussion of finite, undetermined quantum corrections is given
in~\cite{ref-jackiw3}. When a symmetry or other formal property forbids the
appearance of a given operator
at tree level, finite radiative contributions to this operator have been found to
give definite values. For example, the anomalous magnetic moment of the electron
in QED and the photon mass in the Schwinger model~\cite{ref-schwinger} get
definite values from loop corrections. Of these two quantities, the
former is forbidden at tree
level by the requirement
of renormalizability and the latter by gauge invariance. However, the photon mass
in the chiral Schwinger model~\cite{ref-jackiw4} is undetermined, because gauge
invariance cannot be preserved in that theory. Similarly, if Lorentz and CPT
invariances are abandoned, no other symmetry can prevent a Chern-Simons term
from being present in the bare action; so,
as discussed above, the finite radiative corrections to this term are
undetermined.

In the theory considered here, both of the regimes discussed
in~\cite{ref-jackiw3} actually manifest themselves, in a self-consistent fashion.
If the gauge symmetry is not spontaneously broken, then the Lorentz-violating
radiative
corrections are uniquely determined; they vanish. On the other hand, if there is
no gauge invariance to protect the theory, gauge-dependent corrections arise.
This situation is not exactly the same as in the other cases discussed above,
because
the parameter describing the ambiguity, $\xi$, is the gauge parameter itself. So
the observations
that the corrections are undetermined and that they depend on the gauge
both come directly from the fact that there is a nontrivial $\xi$ dependence;
in other words, the gauge dependence is the ambiguity.

In this framework of the preceding paragraph, the
Lorenz-Landau gauge manifests itself as a fine tuning of
the model, such that the Lorentz-violating effects are made to vanish. However,
if it actually turns out that the theory is only renormalizable and unitary in
this particular gauge, then these conditions will again restore uniqueness to
the radiative corrections. A well-definied theory can result only if $\xi=0$, so
the only radiative corrections that could be seen physically would correspond to
this special value.

\section{Conclusion}
\label{sec-conclusion}

We have demonstrated that there are subtle issues surrounding any
Lorentz-violating operators that involve Faddeev-Popov ghost fields.
We introduced a Lorentz-violating
term into the ghost sector of the action. In addition to Lorentz symmetry, we
expected this addition to break the gauge symmetry of the theory, if it actually
turned out to have any
physical effects. However, the question of whether such physical effects
exist turns out to be quite subtle.

It is well
established that some Lorentz-violating operators have no physical consequences.
However, in an Abelian gauge theory, one cannot determine whether or not the
$c^{\nu\mu}$ operator is physically meaningful without knowing
what phase the theory is in. Moreover, when the Lorentz violation does exist, it
is gauge dependent. So far, our results only apply to $U(1)$ gauge theories;
however, it would be very interesting to see how they generalize to the
non-Abelian case, in which the ghosts are more closely coupled to the rest of the
theory.

\section*{Acknowledgments}
The author is grateful to R. Jackiw and V. A. Kosteleck\'{y} for
helpful discussions.
This work is supported in part by funds provided by the U. S.
Department of Energy (D.O.E.) under cooperative research agreement
DE-FG02-91ER40661.


\begin{thebibliography}{99}

\bibitem{ref-kost1}D. Colladay, V. A. Kosteleck\'{y}, Phys. Rev. D {\bf 55},
6760 (1997).
\bibitem{ref-kost2}D. Colladay, V. A. Kosteleck\'{y}, Phys. Rev. D {\bf 58},
116002 (1998).
\bibitem{ref-kost12}V. A. Kosteleck\'{y}, Phys. Rev. D {\bf 69}, 105009 (2004).
\bibitem{ref-kost3}V. A. Kosteleck\'{y}, R. Lehnert, Phys. Rev. D {\bf 63},
065008 (2001).
\bibitem{ref-kost4}V. A. Kosteleck\'{y}, C. D. Lane, A. G. M. Pickering,
Phys. Rev. D {\bf 65}, 056006 (2002).
\bibitem{ref-bluhm1}R. Bluhm, V. A. Kosteleck\'{y}, N. Russell, Phys. Rev.
Lett. {\bf 79}, 1432 (1997).
\bibitem{ref-bluhm2}R. Bluhm, V. A. Kosteleck\'{y}, N. Russell, Phys. Rev. D
{\bf 57}, 3932 (1998).
\bibitem{ref-gabirelse}G. Gabrielse, A. Khabbaz, D. S. Hall, C. Heimann, H.
Kalinowsky, W. Jhe, Phys. Rev. Lett. {\bf 82}, 3198 (1999).
\bibitem{ref-dehmelt1}H. Dehmelt, R. Mittleman, R. S. Van Dyck, Jr., P.
Schwinberg, Phys. Rev. Lett. {\bf 83}, 4694 (1999).
\bibitem{ref-bluhm3}R. Bluhm, V. A. Kosteleck\'{y}, N. Russell , Phys. Rev.
Lett. {\bf 82}, 2254 (1999).
\bibitem{ref-phillips}D. F. Phillips, M. A. Humphrey, E. M. Mattison, R. E.
Stoner, R. F. C. Vessot, R. L. Walsworth , Phys. Rev. D {\bf 63}, 111101 (R)
(2001).
\bibitem{ref-kost8}R. Bluhm, V. A. Kosteleck\'{y}, C. D. Lane, Phys. Rev. Lett.
{\bf 84}, 1098 (2000).
\bibitem{ref-hughes}V. W. Hughes, {\em et al.}, Phys. Rev. Lett. {\bf 87},
111804 (2001).
\bibitem{ref-kost9}R. Bluhm, V. A. Kosteleck\'{y}, Phys. Rev. Lett. {\bf 84},
1381 (2000).
\bibitem{ref-heckel}B. Heckel, {\em et al.}, in {\em Elementary Particles and
Gravitation}, edited by B. N. Kursonoglu, {\em et al.} (Plenum, New York, 1999).
\bibitem{ref-berglund}C. J. Berglund, L. R. Hunter, D. Krause, Jr., E. O.
Prigge, M. S. Ronfeldt, S. K. Lamoreaux, Phys. Rev. Lett. {\bf 75}, 1879 (1995).
\bibitem{ref-kost6}V. A. Kosteleck\'{y}, C. D. Lane, Phys. Rev. D {\bf 60},
116010 (1999).
\bibitem{ref-bear}D. Bear, R. E. Stoner, R. L. Walsworth, V. A. Kosteleck\'{y},
C. D. Lane, Phys. Rev. Lett. {\bf 85}, 5038 (2000).
\bibitem{ref-kost10}V. A. Kosteleck\'{y}, Phys. Rev. Lett. {\bf 80}, 1818
(1998).
\bibitem{ref-kost7}V. A. Kosteleck\'{y}, Phys. Rev. D {\bf 61}, 016002 (2000).
\bibitem{ref-hsiung}Y. B. Hsiung, Nucl. Phys. Proc. Suppl. {\bf 86}, 312
(2000).
\bibitem{ref-abe}K. Abe {\em et al.}, Phys. Rev. Lett. {\bf 86}, 3228 (2001).
\bibitem{ref-carroll1}S. M. Carroll, G. B. Field, R. Jackiw, Phys. Rev. D
{\bf 41}, 1231 (1990).
\bibitem{ref-carroll2}S. M. Carroll, G. B. Field, Phys. Rev. Lett. {\bf 79},
2394 (1997).
\bibitem{ref-kost11}V. A. Kosteleck\'{y}, M. Mewes, Phys. Rev. Lett. {\bf 87},
251304 (2001).
\bibitem{ref-jackiw2}R. Jackiw, S. Templeton, Phys. Rev. D {\bf 23}, 2291
(1981).
\bibitem{ref-schonfeld}J. Schonfeld, Nucl. Phys. B {\bf 185}, 157 (1981).
\bibitem{ref-coleman1}S. Coleman, S. L. Glashow, Phys. Rev. D {\bf 59}, 116008
(1999).
\bibitem{ref-jackiw1}R. Jackiw, V. A. Kosteleck\'{y}, Phys. Rev. Lett.
{\bf 82}, 3572 (1999).
\bibitem{ref-victoria1}M. P\'{e}rez-Victoria, Phys. Rev. Lett. {\bf 83}, 2518
(1999).
\bibitem{ref-chen}W. F. Chen, Phys. Rev. D {\bf 60}, 085007 (1999).
\bibitem{ref-chung3}J. M. Chung, Phys. Rev. D {\bf 60}, 127901 (1999).
\bibitem{ref-altschul1}B. Altschul, Phys. Rev. D {\bf 69}, 125009 (2004).
\bibitem{ref-altschul2}B. Altschul, Phys. Rev. D {\bf 70}, 101701 (2004).
\bibitem{ref-coleman2}S. Coleman, E. Weinberg, Phys. Rev. D {\bf 7}, 1888 (1973).
\bibitem{ref-feddeev}L. D. Faddeev, V. N. Popov, Phys. Lett. B {\bf 25}, 29
(1967).
\bibitem{ref-becchi}C. Becchi, A. Rouet, R. Stora, Ann. Phys. {\bf 98}, 287
(1976).
\bibitem{ref-tyutin}M. Z. Iofa, I. V. Tyutin, Theor. Math. Phys. {\bf 27}, 316
(1976).
\bibitem{ref-colladay2}D. Colladay, P. McDonald, J. Math. Phys. {\bf 43}, 3554
(2002).
\bibitem{ref-altschul3}B. Altschul, D. Colladay, Phys. Rev. D {\bf 71}, 125015
(2005).
\bibitem{ref-dolan}L. Dolan, R. Jackiw, Phys. Rev. D {\bf 9}, 2904 (1974).
\bibitem{ref-jackiw3}R. Jackiw, Int. J. Mod. Phys. B {\bf 14}, 2011 (2000).
\bibitem{ref-schwinger}J. Schwinger, Phys. Rev. {\bf 128}, 2425 (1962).
\bibitem{ref-jackiw4}R. Jackiw, R. Rajaraman, Phys. Rev. Lett. {\bf 54}, 1219
(1985).

\end{thebibliography}
\end{document}